\documentstyle[epsfig]{aipproc}

\def\etal{{\it et al.}}

\begin{document}
{\centerline{To appear in the Proc. 4th Huntsville GRB Symp.}}

\title{The Gamma Ray Bursts GRB970228 and GRB970508: What Have We Learnt?}

\author{Mario Livio$^{1}$, Kailash C.~Sahu$^{1}$, Larry Petro$^{1}$,
Andrew Fruchter$^1$, Elena Pian$^2$, F.~Duccio Macchetto$^{1}$, 
Jan van Paradijs$^{3,4}$, Chryssa Kouveliotou$^5$, Paul J.~Groot$^{3}$,
Titus J. Galama$^3$}

\address{$^1$Space Telescope Science Institute, 
3700 San Martin Drive, Baltimore, MD 21218, USA\\
{$^2$Instituto di Technologie e Studio delle Radiazioni Extraterrestri,
C.N.R., Via Gobetti 101, I-40129 Bologna, Italy}\\
{$^3$Astronomical Institute ``Anton Pannekoek", 
University of Amsterdam, \& Center for High Energy Astrophysics, 
Kruislaan 403, 1098 SJ Amsterdam, The Netherlands}\\
{$^4$Physics Department,
University of Alabama in Huntsville, Huntsville, AL 35899, USA}\\
{$^5$Universities Space Research Association, 
NASA Marshall Space Flight Center, ES-84, Huntsville, AL 35812, USA}\\
}
\maketitle
\begin{abstract}
We examine what we regard as key observational results on GRB 970228 and 
GRB 970508 and show that the accumulated evidence strongly suggests
that $\gamma$-ray bursts (GRBs) are cosmological fireballs.

We further show that the observations suggest that GRBs are not 
associated with the nuclear activity of active galactic
nuclei, and that late-type galaxies are more prolific producers of GRBs.

We suggest that GRBs can be used to trace the cosmic history of
the star-formation rate. Finally, we show that the GRB locations 
with respect to
the star-forming regions in their host galaxies and the total 
burst energies can be used to distinguish between different theoretical models 
for GRBs.

\end{abstract}

\section*{Introduction}

It is very often the case in astronomy that multiwavelength observations of a single 
object allow a dramatic progress in the understanding of the object, and $\gamma$-ray 
bursts (see e.g. Fishman and Meegan, 1995) proved to be no exception. The identification of the X-ray and optical
counterparts of the two $\gamma$-ray burst sources GRB 970228 and GRB 
970508 marks a remarkable milestone in the research of these enigmatic objects
(e.g. Costa et al. 1997 a,b; Heise et al. 1997; Piro et al. 1997;
van Paradijs et al. 1997; Sahu et al. 1997 a,b; Bond 1997; Djorgovski et al. 1997; 
and references therein). In the present short note, we
first present what we regard as the key findings and
their potential implications, and we then examine possibilities to
make further progress with regard to specific $\gamma$-ray burst models.

\section*{Key Observational Results and Their Main Implications }

We regard the following observational findings (in no
particular order) as providing the main 
clues for the understanding of the nature of $\gamma$-ray bursts:

(1) Contrary to claims made by Caraveo et al. (1997 a,b), {\it no
significant proper motion has been detected for GRB 970228}
(the limit is 36 milliarcsec per year, Sahu et al. 1997 a,b; Fruchter et al. 1998). This makes an
extremely close origin for the GRB very unlikely.

(2){\it  An absorption- and emission-line system at z=0.835 has been 
detected in the
optical spectrum of GRB 970508} (Metzger et al. 1997 a,b).

Since this absorption/emission system probably arises from a host (or intervening)
galaxy, this provides direct evidence that this GRB is at a cosmological distance 
(assuming, of course, that the identification of the source is correct).

(3) The principal features of the afterglow in the optical (see Fig. 1)
are well represented by a forward-radiating blast-wave model 
(M\'esz\'aros and Rees 1997).
In fact, given the fact that the afterglows may be expected to depend 
on the properties of the environment of the GRB and on the angular anisotropy
of the fireball itself (e.g. M\'esz\'aros, Rees and Wijers 1997),
the agreement with the simplest model can be considered
quite remarkable.

\begin{figure}
\centerline{\epsfig{file=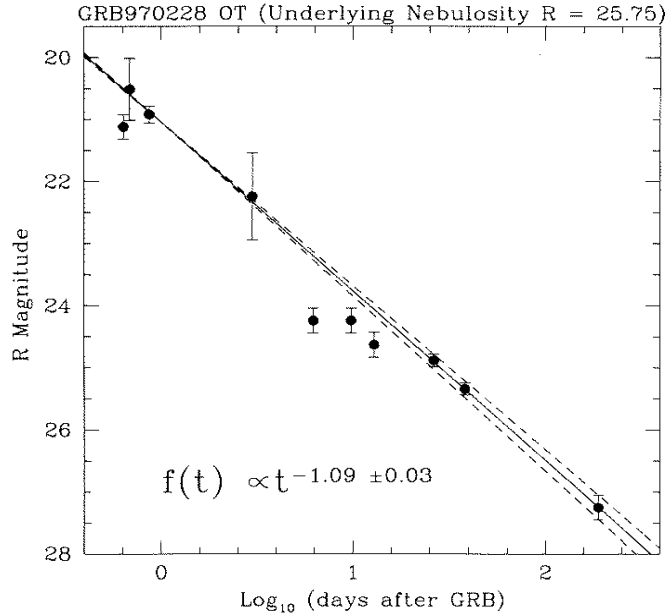,height=3.5in,width=3.5in}}
\vspace{10pt}
\caption{The decay behavior of the optical transient 
associated with GRB 970228 in R-band.}
\label{fig1}
\end{figure}

(4) {\it No break  towards a more rapid 
decline has been seen so far in the power-law behavior of the 
light curve (Fig. 1).} Such a break is expected to occur 
after the blastwave has `snowploughed' through a rest-mass energy of the
order of the burst energy, since the remnant then becomes nonrelativistic (e.g.
Wijers, Rees and M\'esz\'aros 1997).
The timescale for such a break to occur is given by

\begin{equation}
 t_{break} \simeq \left( {3E_\gamma\over{4\pi \rho c^5}}\right) ^{1/3} =
\left\{ \begin{array}{c}
1 yr \left( {E_\gamma\over{10^{51} erg}}\right) ^{1/3} \left( 
{n\over{0.1 cm^{-3}}}\right) ^{1/3}\\
2 days \left( {E_\gamma\over{10^{42} erg}}\right) ^{1/3} \left( 
{n\over{0.001 cm^{-3}}}\right) ^{1/3}  \end{array} \right.  
\end{equation}
 
\noindent where $\rho$ is the medium density (n is the number density),
E$_\gamma$ is the burst energy, and we have scaled these quantities 
with values appropriate for a cosmological or extended halo origin, 
respectively. 
Eq. (1) clearly demonstrates that the fact that a break has not yet been 
observed in GRB 970228, more than six months after the burst,  
strongly favors a cosmological origin.

(5) {\it A potential host galaxy has been identified in the case of GRB 970228 }
(Sahu et al. 1997a; Fruchter et al. 1998; see Fig. 2). The probability of a chance 
superposition of the source with a galaxy of that magnitude (V $\sim$ 25.7)
is of the order of 2\% (Fruchter et al. 1997a).

An examination of points (1) -(5) above clearly suggests that GRBs are 
cosmological fireballs. While it is certainly true, that with only
two optical afterglows observed so far, an alternative explanation can
be found for each one of the above points 
(e.g. the host galaxy of GRB 970228 could be a chance superposition afterall;
what is thought to be the counterpart of GRB 970508 may be an unrelated BL Lac object, etc.),
the combined weight of all the observational facts strongly argues against a local origin
for the GRBs. We will therefore from here on {\it assume} that GRBs are cosmological, and that
GRB 970228 is indeed located in what appears to be its host galaxy.
 
\section*{General Implications and Specific Models for GRBs}

A close examination of the HST images of GRB 970228 (Sahu et al. 1997a;
Fruchter et al. 1997a) reveals two more observational
facts:

(i) The GRB is not at the center of the host galaxy.

(ii) The galaxy looks like a dwarf irregular or spiral, not like an elliptical
galaxy.

The first of these facts implies that {\it GRBs are probably not
associated with the central massive black holes
in their host galaxies}, or in general, with the
nuclear activity of active galactic nuclei.
This rules out, for example, tidal disruption of stars (Carter, 1992)
as potential models for GRBs. The second point is somewhat less
certain, because the faintness of the galaxy makes any morphological
determination not entirely conclusive. Nevertheless, taken at face value,
this observation suggests that the frequency of GRBs may be {\it higher
in late-type galaxies}. This, in turn, favors models for GRBs 
which involve a young stellar population. 
Leading models in this category include: merging neutron stars (e.g. Eichler et al. 1989)
or a neutron star and a black hole (e.g. Mochkovitch et al. 1993),
``failed type Ib supernovae" (Woosley 1993), ``hypernovae" (Paczy\'nski 1997a),
radio pulsar glitches (Melia and Fatuzzo 1992) and, to a
lesser extent, collapsing white dwarfs (Usov 1992). 
For example, Sahu et al (1997a) have shown that the ratio of the
neutron-star merger rate in disk galaxies to that in ellipticals
is about 80 (see also Phinney 1991; Narayan, Piran and Shemi 1991).

\begin{figure}
\centerline{\epsfig{file=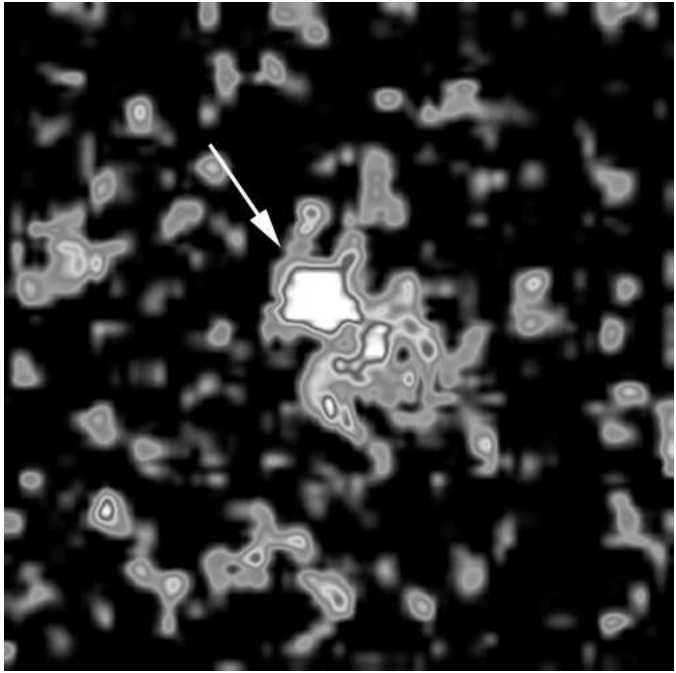,height=3.5in,width=3.5in}}
\vspace{10pt}
\caption{Smoothed HST image of GRB 970228, taken in March/April 1997 with 
the WFPC2 camera (Sahu et al. 1997a).}
\label{fig2}
\end{figure}

Another important thing to realize is the fact that {\it all} the leading models for GRBs have
relatively short delays ($\le 10^9$ yr) with respect to the
star formation process. Consequently, {\it GRBs essentially trace the redshift
distribution of the star-formation rate} (Sahu et al. 1997b;
Totani 1997; Wijers et al. 1997). This means that one can use the inferred cosmic history of the star-formation rate (Madau et al.
1996; Connolly et al. 1997) to construct a synthetic log N - log P relation for GRBs. Turned around,
given enough statistics, the redshift distribution
of GRBs can provide {\it an independent test} for the cosmic history of the
star formation rate.

Finally, we can ask: is the available data sufficient to point us 
towards a specific model for the GRBs?

\begin{table}
\caption{Locations of GRBs and Total Energies in Different Models.}
\label{table1}
\begin{tabular}{llcc}
Model&Location of GRB & Total Energy (ergs)& \cr
\tableline
Failed supernova& In star-forming region& $\sim$10$^{51}$& \cr
Hypernova& In star-forming region& $\sim$10$^{54}$& \cr
White-dwarf collapse& In disks or globular clusters& $\sim$10$^{51}$& \cr
Merging neutron stars or & Up to tens of kpc away & & \cr
neutron star+BH& from the star-forming region& $\sim$10$^{51}$& \cr
\end{tabular}
\end{table}

There are at present two pieces of information which can, in principle,
provide clues in this direction. One is related to the burst energy and 
the other to the burst location (see also Pazy\'nski 1997b).
In Table 1, we list some of the most popular GRB models, the total energy
expected from these models, and the expected location of the GRBs with 
respect to the birth place of their progenitors. 
One should note that failed supernovae (Woosley 1993), 
hypernovae (Paczy\'nski 1997a) and white dwarf collapses (Usov 1992) are all
expected to be found near the birth place of their progenitors for the
following reasons. Since failed supernovae and hypernovae originate from
very massive stars, the lifetime of their progenitors is short, $\sim$
10$^6$ yrs, and consequently, they cannot travel very far. 
The progenitor lifetime of a white dwarf collapse 
can be longer, but its space velocity is typically low. 
On the other hand, double neutron star systems are expected to be born with high 
kick velocities of the order of a few hundred km s$^{-1}$ (Lyne and
Lorimer 1994; White and van Paradijs 1996; Fryer and Kalogera 1997), 
and it takes them
typically a long time ($\sim 10^8$ yr) to merge. Consequently,
the GRBs in this case can be expected to be found
typically 30 kpc away from their birthplace. 
Unfortunately, the existing data do not provide yet a clear picture
concerning the burst locations. For example, while the
detection of the [OII] 3728\AA \ emission line (Metzger et al 1997b)
in GRB 970508 seems to indicate the presence of a relatively dense
medium (and thus, a potential association with a star-forming region),
HST failed so far to detect a host galaxy at the burst location
(Fruchter et al. 1997b), while two nearby
faint galaxies (which could, in principle, be the hosts) were
detected at distances of $\sim$30 kpc and 35 kpc (for z=0.835, H$_0$ 
= 65 km s$^{-1}$ Mpc$^{-1}$). Similarly, the location of GRB 970228 near the
edge of its host galaxy does not give us conclusive information on whether
the progenitor has moved from its birth place or not.

Concerning the total energy, again the present data are still very 
inconclusive. For example, the total energy inferred for GRB 970508 was $\sim$ 10$^{52}$
ergs, assuming spherical emission (Waxman 1997a,b). The situation is further complicated
by the possibility that the $\gamma$-ray emission is beamed (e.g. Wijers et al. 1997b).

In spite of the difficulties pointed out above, the burst locations and the total energies
do hold the potential of providing in the future the next step
from generic fireballs to specific physical models.


\begin{references}
\bibitem{} Bond, H.E., 1997, IAU Circ. No. 6654.
\bibitem{} Caraveo, P.A., Mignami, R., Tavani, M., and Bignami, G.F., 1997a, IAU 
Circ. No. 6629.
\bibitem{} Caraveo, P.A., Mignami, R., Tavani, M., and Bignami, G.F., 1997b, 
Astron. Astrophys, 326, l13
\bibitem{} Carter, B. 1992, ApJ. 391, L67.
\bibitem{} Connolly, A.J., Szalay, A.S., Dickinson, M., Subbarao, M.V.,
Brunner, R.J., 1997, ApJ., 486, L11.
\bibitem{} Costa, E. \etal, 1997a, preprint astro-ph/9706065.
\bibitem{} Costa, E. \etal, 1997b, IAU Circ. No. 6572.
\bibitem{} Djorgovski, S.G. \etal, 1997, Nature, 387, 876
\bibitem{} Eichler, D., Livio, M., Piran, T., and Schramm, D.N., 
1989, Nature, 340, 126.
\bibitem{}  Fishman, G.J., \& Meegan, C.A. 1995, Ann. Rev.
Astron. Astrophys. 33, 415.
\bibitem{} Fruchter, A. \etal, 1997a, ApJ, submitted.
\bibitem{} Fruchter, A. \etal, 1997b, IAU Circ. No. 6674.
\bibitem{} Fryer, C. and Kalogera, V., 1997, ApJ, in press.
\bibitem{} Heise, J. \etal, 1997, IAU Circ. No. 6654.
\bibitem{} Lyne, A.G. and Lorimer, D.R., 1994, Nature, 369, 127.
\bibitem{} Madau, P. \etal, 1996, MNRAS, 283, 1388.
\bibitem{} Melia, F., Fatuzzo, M., 1992, ApJ., 398, L85.
\bibitem{} M\'esz\'aros, P., and Rees, M.J., 1997, ApJ, 476, 232.
\bibitem{} M\'esz\'aros, P., Rees, M.J., Wijers, R.A.M., 1997, 
astro-ph/9704153
\bibitem{} Metzger, M.R. \etal, 1997a, IAU Circ. No. 6676.
\bibitem{} Metzger, M.R. \etal, 1997b, Nature, 387, 879
\bibitem{} Mochkovitch, R., Hernanz, M., Isern, J., Martin, X., 1993,
Nature, 361, 236
\bibitem{} Narayan, R., Paczy\'nski, B., and Piran, T., 1992, ApJ, 
395, L83.
\bibitem{} Narayan, R., Piran, T., Shemi, A., 1991, ApJ, 379, L17.
\bibitem{} Paczy\'nski, B., 1997a, astro-ph/9706232
\bibitem{} Paczy\'nski, B., 1997b, astro-ph/9710086
\bibitem{} van Paradijs, J. \etal, 1997, Nature, 386, 686.
\bibitem{} Phinney, E.S. 1991, ApJ.  380, L17.
\bibitem{} Piro, L., \etal, 1997, IAU Circ. No. 6656.
\bibitem{} Sahu, K.C. \etal, 1997a, Nature, 387, 476.
\bibitem{} Sahu, K.C. \etal, 1997b, ApJL, 489, L127.
\bibitem{} Totani, T., 1997, ApJ., 486, 71
\bibitem{} Usov, V., 1992, Nature, 357, 472
\bibitem{} Waxman, E., 1997a, ApJ, 489, L33
\bibitem{} Waxman, E., 1997b, ApJ, 485, L5
\bibitem{} White, N.E., and van Paradijs, J., 1996, ApJ, 473, L25.
\bibitem{} Wijers, R.A.M.J., Rees, M.J., and M\'esz\'aros, P., 1997, 
MNRAS, 288, L51.
\bibitem{} Wijers, R.A.M.J., Bloom, J.S., Bagla, J.S.,
Natarajan, P., 1997, astro-ph/9708183.
\bibitem{} Woosley, S., 1993, ApJ., 405, 273.
\end{references}
\end{document}